\documentclass[aip,apl,reprint,a4paper]{revtex4-2}
\usepackage[utf8x]{inputenc}
\usepackage{graphicx}
\usepackage{textcomp}
\usepackage{fourier}
\usepackage{censor}
\usepackage[american]{babel}

\begin{document}

\title{Adsorption-controlled epitaxy of perovskites}
\author{Wolfgang Braun}
\affiliation{Max Planck Institute for Solid State Research, Heisenbergstr.\ 1, 70569 Stuttgart, Germany}
\date{May 2, 2018}

\begin{abstract}
I propose to use laser heating both for the substrate and the thermal evaporation sources in a vacuum chamber operating at pressures from XHV to values where the mean free path of the particles approaches or slightly exceeds the source--substrate distance.
The concept combines the advantages of the molecular beam epitaxy (MBE) and pulsed laser deposition (PLD) methods to allow ultrapure deposition with continuous stoichiometry variation at high background pressures of arbitrary gases or molecular beams.
Theory and preliminary experiments suggest that this setup is capable of growing complex oxides such as SrTiO$_3$ in the adsorption-controlled regime, similar to GaAs, in a background of molecular oxygen.
This regime is neither accessible to MBE nor to PLD, making this  laser epitaxy approach a unique tool to explore new growth regimes with the potential to fabricate structures such as modulation-doped heterostructures with low levels of background impurities that are impossible to synthesize with the current techniques.
The technological simplicity and exceedingly compact size of the deposition chamber enable easy and rapid switching between different materials systems and the efficient synthesis of new materials that involve corrosive constituents.
In contrast to PLD, the method may be scaled in a straightforward manner to large substrate sizes, providing a direct path from research to mass production.
\end{abstract}

\maketitle

\section{Problem}
In the ideal case of molecular beam epitaxy (MBE), the epitaxial layer on the substrate is formed from thermally evaporated beams of the individual elements that synthesize the desired compound at the substrate surface.
Ideally, the growth is performed in the so-called adsorption-controlled regime, where one or more of the constituents of the compound are only incorporated into the growth front if the other constituent(s) are present in the correct stoichiometric ratio.
Superfluous material desorbs from the surface.
This allows the self-adjusting growth of the exact stoichiometry by offering a surplus of the desorbing species, allowing the non-desorbing species to always find a reaction partner to form the stoichiometric compound.
Growth is then controlled by the supply of the non-desorbing species, called the rate-limiting species.
The traditional example for this growth mode is the MBE of GaAs or AlAs, where As desorbs in a wide range of growth conditions  if no matching Ga or Al is present.
This wide \lq growth window\rq\ has been essential for the success of III-V epitaxy and is a desired feature for any vacuum crystal growth method.

Traditionally, both the substrate and the sources in MBE are heated by resistive heaters, their temperature being controlled with thermocouple sensors.
This results in good stability, precision and reproducibility of the available fluxes, and therefore good composition control.
The use of ultrapure elemental sources and the ability to bake these prior to deposition, results in high purity of the deposited layers and therefore high structural perfection.
On the other hand, it introduces a significant number of hot filaments and electrical feedthroughs to the vacuum system, posing challenges for the uptime and productivity of the method.
Filaments are also sensitive to corrosive gases such as oxygen, Te or S, which are desirable for the growth of oxides, phase-change materials and 2D layered compounds.
The background pressures of such gases are therefore usually limited to the 10$^{-6}$\,mbar range.

\noindent MBE pro:
\begin{itemize}
\item precise and gradually adjustable composition and dopant control
\item high purity because of elemental sources that can be baked individually
\end{itemize}
MBE con:
\begin{itemize}
\item technologically complex due to many complex components in the vacuum chamber
\item limited to low background pressures, maximum $\approx$1$\times$10$^{-5}$\,mbar
\item only background gases with low corrosivity possible to avoid corrosion of in-vacuum components
\end{itemize}

Pulsed laser deposition (PLD), on the other hand, is conceptually very simple and straightforward.
Inside the process chamber, one needs only a mechanical holder for the target and a mechanical holder for the substrate.
Substrate heating and material ablation are both achieved by laser beams that are generated outside the process chamber.
This allows the use of practically any process gas at almost any pressure, given that the ablated atoms and molecules can still reach the substrate.
It also offers the promise of ultrapure deposition, as the active (sample and target) surfaces are the hottest parts inside the chamber.
There are no other hot surfaces that could generate impurities to condense on the target or substrate surface.
Nor are there sources of energy inside the chamber, since the lasers are outside the vacuum.
The uptime and stability of such a system are therefore potentially very high.

In PLD, the material to be deposited on the substrate is ablated from a target close in stoichiometry to the desired layer composition.
Since the ablation process happens on such short time scales (pulses are 25\,ns wide), a layer of a certain thickness is released from the target quasi instantaneously, and the entire material is ejected towards the substrate where it deposits with essentially the same stoichiometry.
This helpful for the synthesis of multernary compounds.
On the other hand, it precludes stoichiometry variations or gradual dopant profiles that would greatly enlarge the parameter space for the synthesis of functional devices.

\noindent PLD pro:
\begin{itemize}
\item simple setup inside the  vacuum chamber
\item all energy sources outside the vacuum chamber
\item wide range of background pressures, maximum $\approx$1\,mbar
\item almost any kind of background gas possible, including corrosive species such as oxygen
\end{itemize}
PLD con:
\begin{itemize}
\item poor stoichiometry control, no gradual composition variation possible
\item high particle energies at low pressures
\end{itemize}

\subsection{Shifting the Pressure Limit}
\begin{figure}[ht]
\begin{center}
\includegraphics[width=\columnwidth]{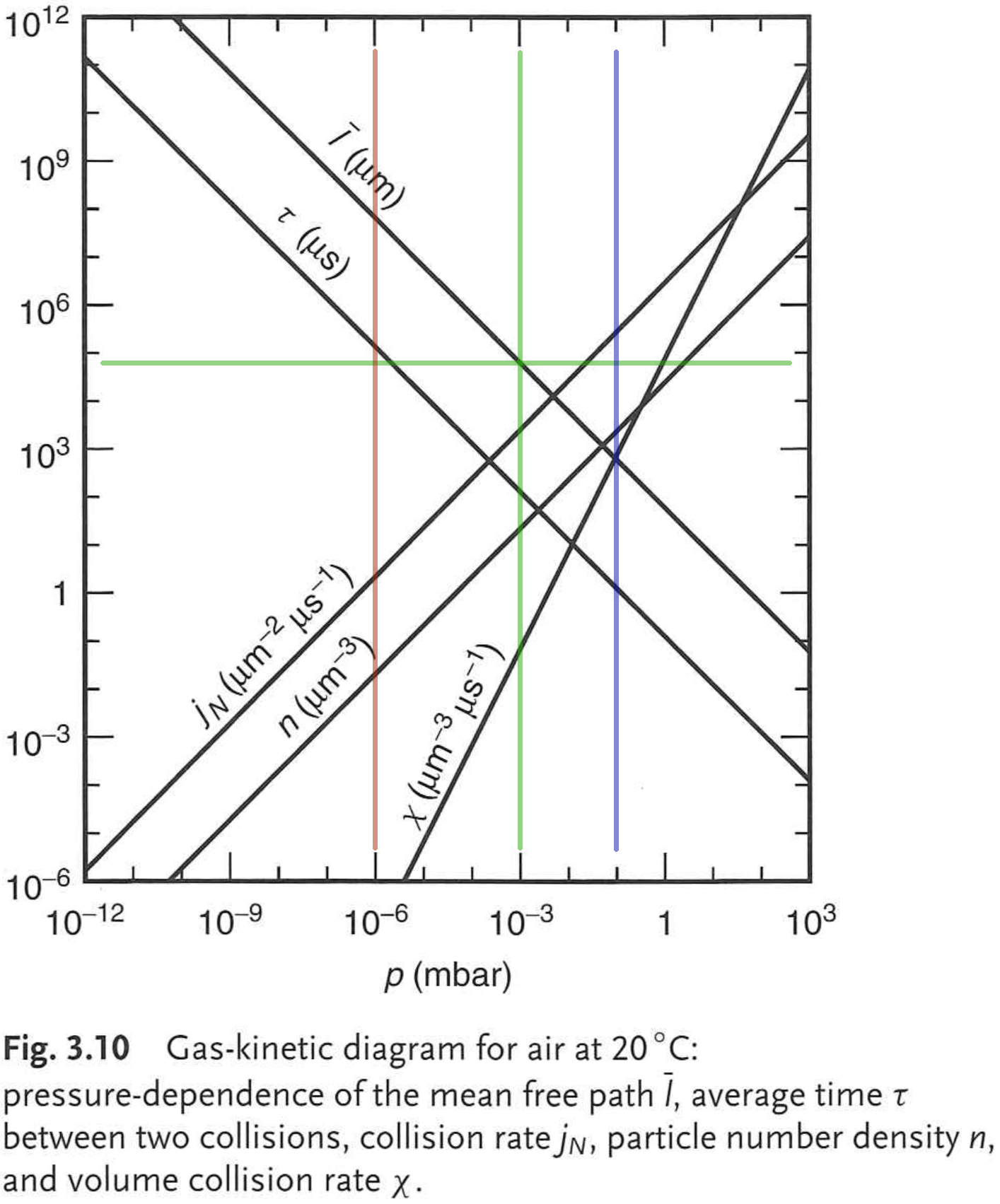}
\end{center}
\caption{Working pressure limits in MBE (red), PLD (blue), and the new method (green).
Adapted from Ref.~\onlinecite{Jousten2008}\label{labelfig_meanfreepath}}
\end{figure}

Figure~\ref{labelfig_meanfreepath} compares the situation for MBE, PLD, and the proposed laser epitaxy concept.
%++++++++++++++++++++++++++++++++++++++++++++++++++++++++++
%++++++++++++++++++++++++++++++++++++++++++++++++++++++++++
We compare the mean free path of background gas, e.g.\ molecular oxygen, with the limits of the growth techniques.
The vertical red line at 10$^{-6}$ mbar indicates the background pressure limit of oxide MBE due to corrosion of filament heaters.
The blue vertical line at 10$^{-1}$ mbar is a typical value for PLD with a working distance of 60\,mm, such that the particles in the plume can still reach the substrate before being scattered in the gas phase to form clusters.
If we now assume the same working distance of 60\,mm for a thermal evaporation scheme, we notice that the atoms can still reach the substrate for pressures of 10$^{-3}$ mbar, where the mean free path matches the source-substrate (working) distance.

This is a gain of three orders of magnitude in background (e.g.\ oxygen) pressure compared to MBE.
These three orders of magnitude offer additional headroom for a larger range of growth conditions.

\subsection{Adsorption-Controlled Epitaxy of Oxides}
\begin{figure}[ht]
\begin{center}
\includegraphics[width=\columnwidth]{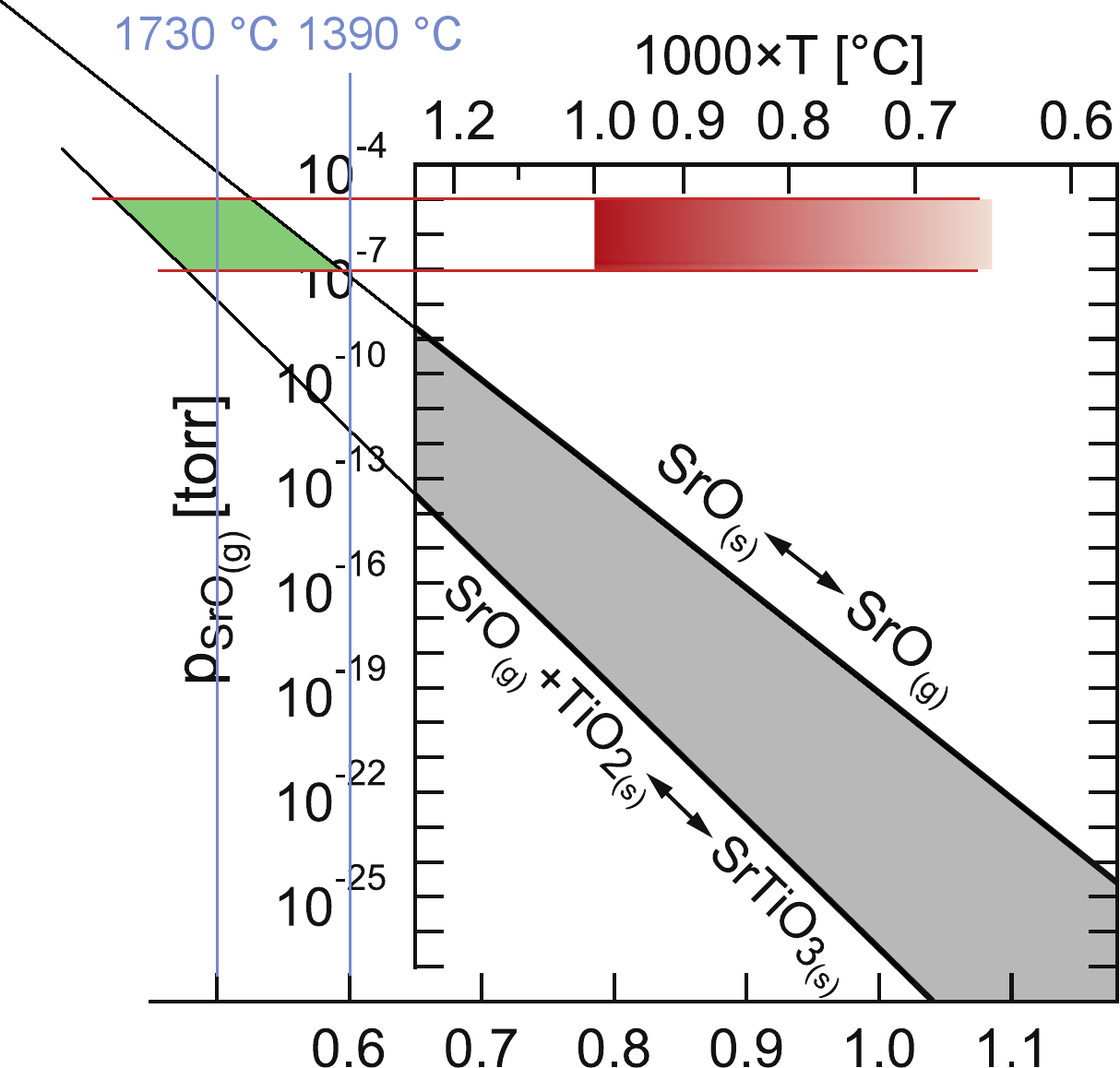}
\end{center}
\caption{Adsorption-limited growth conditions for SrTiO$_3$.
Adapted from Ref.~\onlinecite{EngelHerbert2013417}\label{labelfig_stolimited}}
\end{figure}

Strontium Titanate, SrTiO$_3$, has been studied in detail for growth in MBE, as it is a common oxide substrate material, and also offers interesting physical properties.
It has been found, however, that the adsorption-limited growth conditions typical for III-V epitaxy cannot be reached with current growth equipment.
The situation is depicted in Fig.~\ref{labelfig_stolimited}.
%++++++++++++++++++++++++++++++++++++++++++++++++++++++++++
%++++++++++++++++++++++++++++++++++++++++++++++++++++++++++
In an Arrhenius plot of SrO pressure vs.\ reciprocal temperature, the gray area shows the region of self-adjusting stoichiometry (adsorption-limited growth).
For the entire area covered by the diagram, Ti oxidizes readily, so that the stability of the Ti-O bond does not need to be taken into account.
The lower limit of the gray region is established by the possibility of SrO and TiO$_2$ reacting to SrTiO$_3$.
Above this line, SrTiO$_3$ forms.
The upper limit is defined by the loss of SrO to the gas phase, leading to the decomposition of the SrTiO$_3$.
The graded red bar indicates growth conditions achievable with traditional MBE.
Its vertical width represents typical growth rates between 10 and 1000\,Å/s.
Its horizontal width corresponds to achievable growth temperatures of 1000\,°C with traditional substrate heaters.

\begin{figure}[ht]
\begin{center}
\includegraphics[width=\columnwidth]{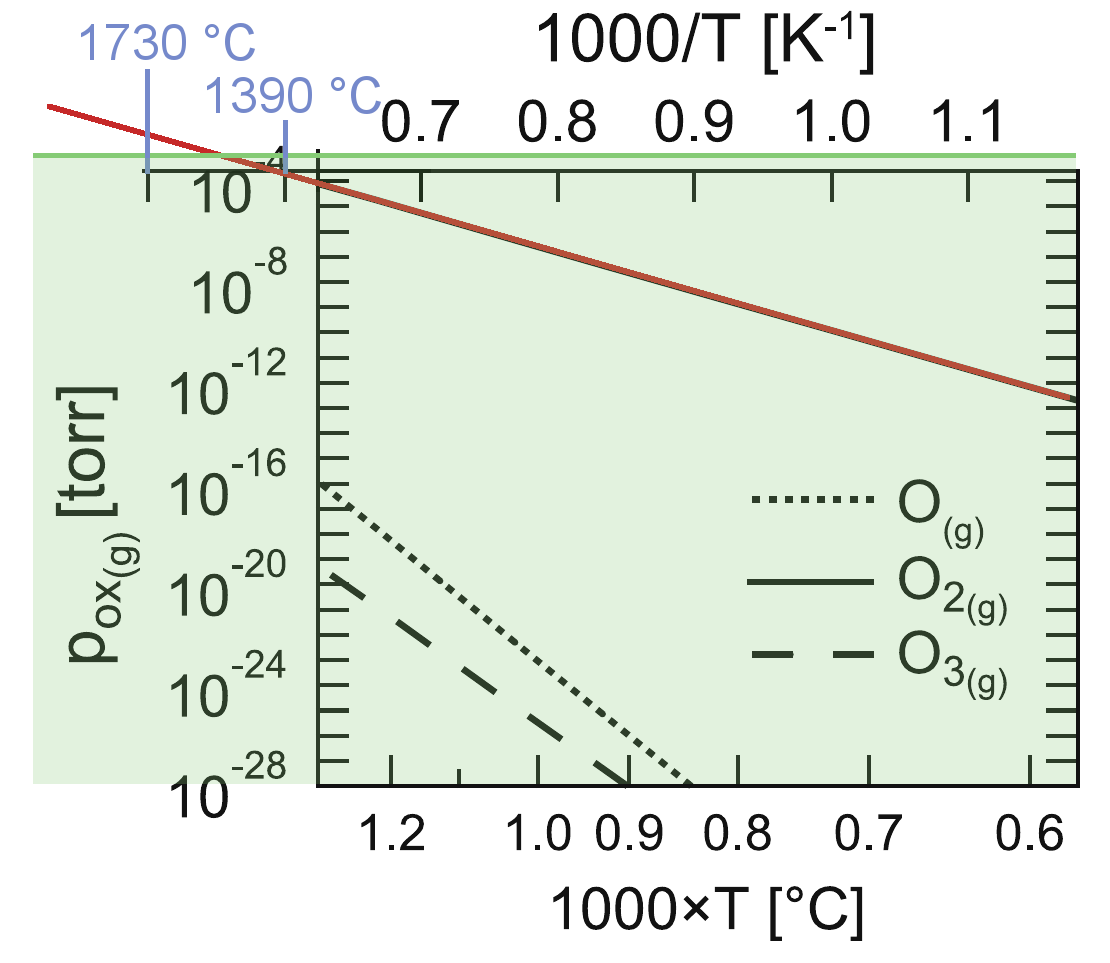}
\end{center}
\caption{Oxidation conditions for SrO.
Adapted from Ref.~\onlinecite{EngelHerbert2013417}\label{labelfig_oxidation}}
\end{figure}

To reach the green area of self-limited adsorption-controlled growth, significantly higher substrate temperatures above 1390\,°C are required.
At the same time, we need to be able to fully oxidize the compound.
The oxidation of the most volatile SrO as a function of temperature and oxygen pressure is shown in Fig.~\ref{labelfig_oxidation}.
%++++++++++++++++++++++++++++++++++++++++++++++++++++++++++
%++++++++++++++++++++++++++++++++++++++++++++++++++++++++++
The calculations are done for atomic/molecular species.
Above the dashed line, Sr forms SrO in the presence of O$_3$.
Above the dotted line, Sr oxidizes in the presence of atomic oxygen.
And finally, Sr oxidizes even with the least reactive molecular oxygen above the solid line.
The range of accessible oxygen pressures in the proposed setup of up to 10$^{-3}$\,mbar is represented by the light green overlay.
The lines intersect at a temperature of approximately 1500\,°C.

\begin{figure}[ht]
\begin{center}
\includegraphics[width=\columnwidth]{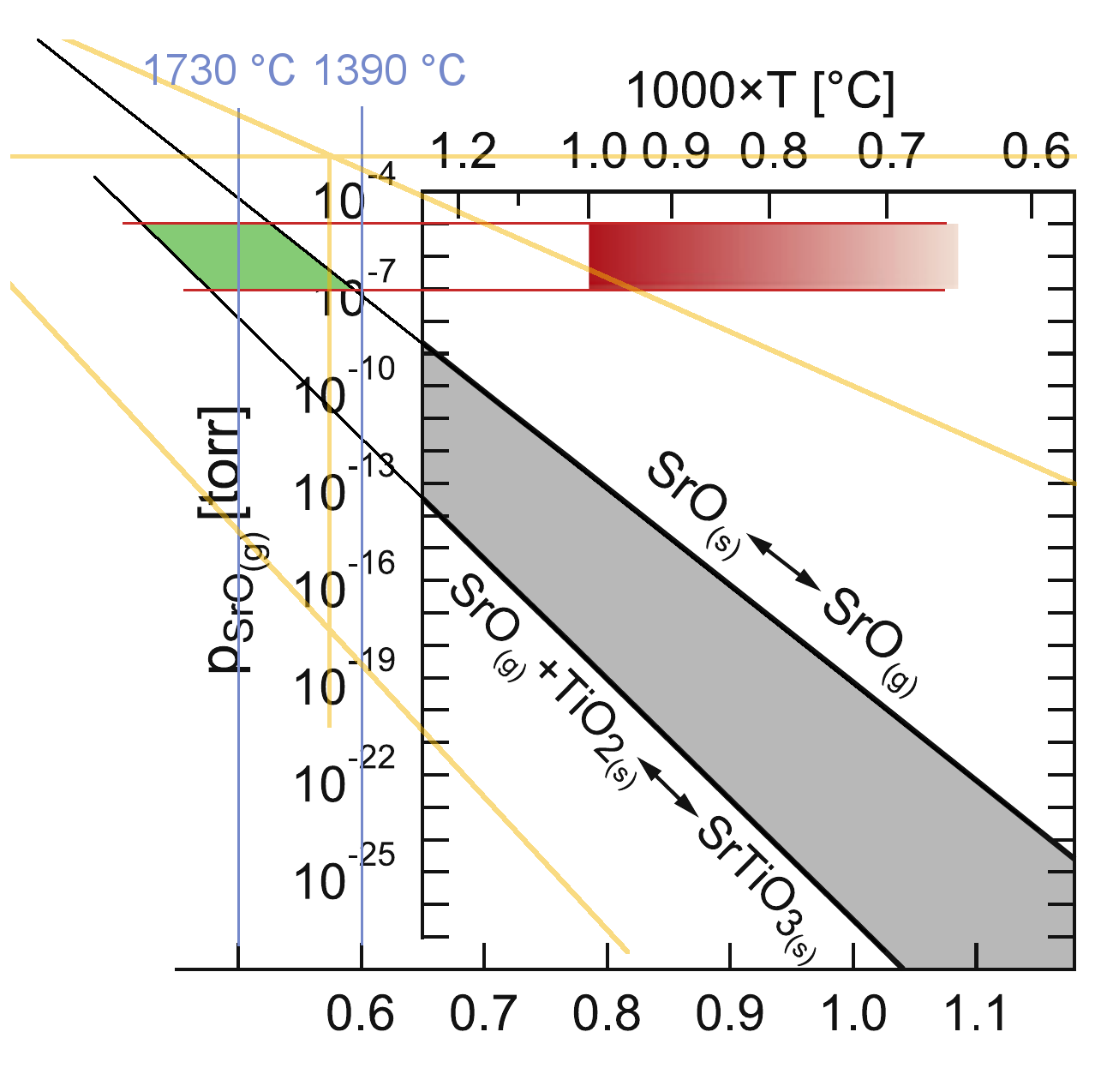}
\end{center}
\caption{Combination of Figs.~\ref{labelfig_stolimited} and \ref{labelfig_oxidation}.
Adapted from Ref.~\onlinecite{EngelHerbert2013417}\label{labelfig_oxidation_plus}}
\end{figure}

Together, both diagrams indicate the existence and accessibility of a self-limited growth regime for SrTiO$_3$ at substrate temperatures above 1300\,°C and molecular oxygen pressures below 10$^{-3}$\,mbar, Fig.~\ref{labelfig_oxidation_plus}.
%++++++++++++++++++++++++++++++++++++++++++++++++++++++++++
%++++++++++++++++++++++++++++++++++++++++++++++++++++++++++
The diagonal orange lines denote the limit of Sr oxidation for molecular oxygen (upper line) and ozone (lower line, copied from Fig.~\ref{labelfig_oxidation}.
At 10$^{-3}$\,mbar of molecular oxygen pressure, the growth window is limited in temperature at around 1400\,°C.
This area is rather small for molecular oxygen, although catalytic cracking of molecular oxygen at the SrTiO$_3$ surface may shift it to lower growth temperatures and/or lower oxygen pressures.
Using ozone at the same pressure, the upper temperature limit shifts to around 2500\,°C, to cover the entire green area shown in Fig.~\ref{labelfig_oxidation_plus}.
The use of inline ozonation, resulting in a few percent of ozone content in the background gas, should therefore allow for complete oxidation.

\section{Solution}
I propose a growth setup designed for the adsorption-limited growth of doped SrTiO$_3$ that will allow us to test the validity of the above considerations.
The concept is based on a blend of PLD and MBE features.

\begin{figure}[ht]
\begin{center}
\includegraphics[width=\columnwidth]{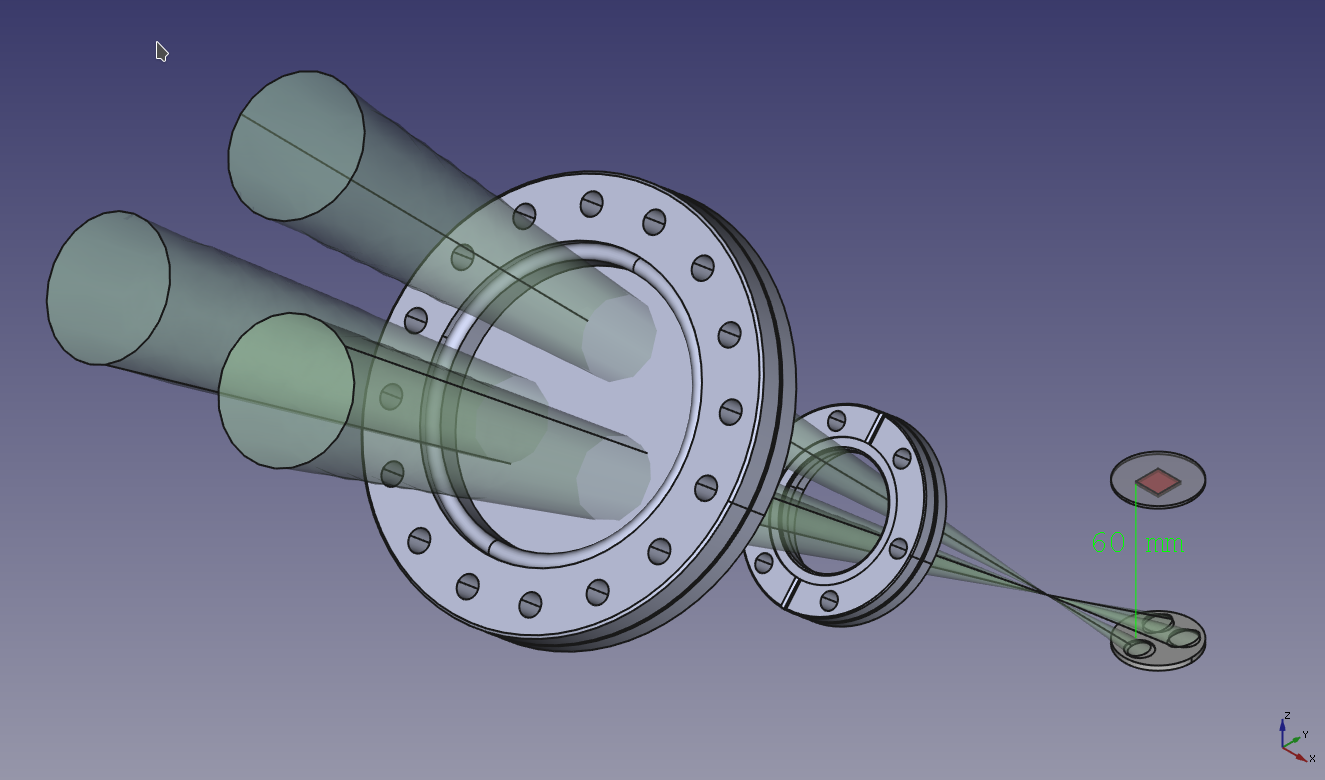}
\end{center}
\caption{Geometry of substrate, source disc and heating laser beams.\label{labelfig_beamtargetsample00}}
\end{figure}

Similar to the geometry of PLD, the sources will be located inside an area 30\,mm in diameter 60\,mm away from the substrate surface, and parallel to it.
Since the elemental source materials can melt, the source plate is located vertically below the substrate, pointing upwards (Fig.~\ref{labelfig_beamtargetsample00}).
%++++++++++++++++++++++++++++++++++++++++++++++++++++++++++
%++++++++++++++++++++++++++++++++++++++++++++++++++++++++++
Similar to a PLD target, the source plate has the same diameter as the substrate holder, and can be transferred in and out of the growth chamber using the same transfer mechanism.
This allows convenient and rapid replenishment of the sources, but also baking of the source plate to high temperatues is possible in a preparation chamber between sample/target introduction and growth chamber for optimum purity.
In its working position, thermocouple temperature sensors may be approached by a linear transfer mechanism to the back of the crucibles, allowing precise temperature control similar to MBE sources.

\begin{figure}[ht]
\begin{center}
\includegraphics[width=\columnwidth]{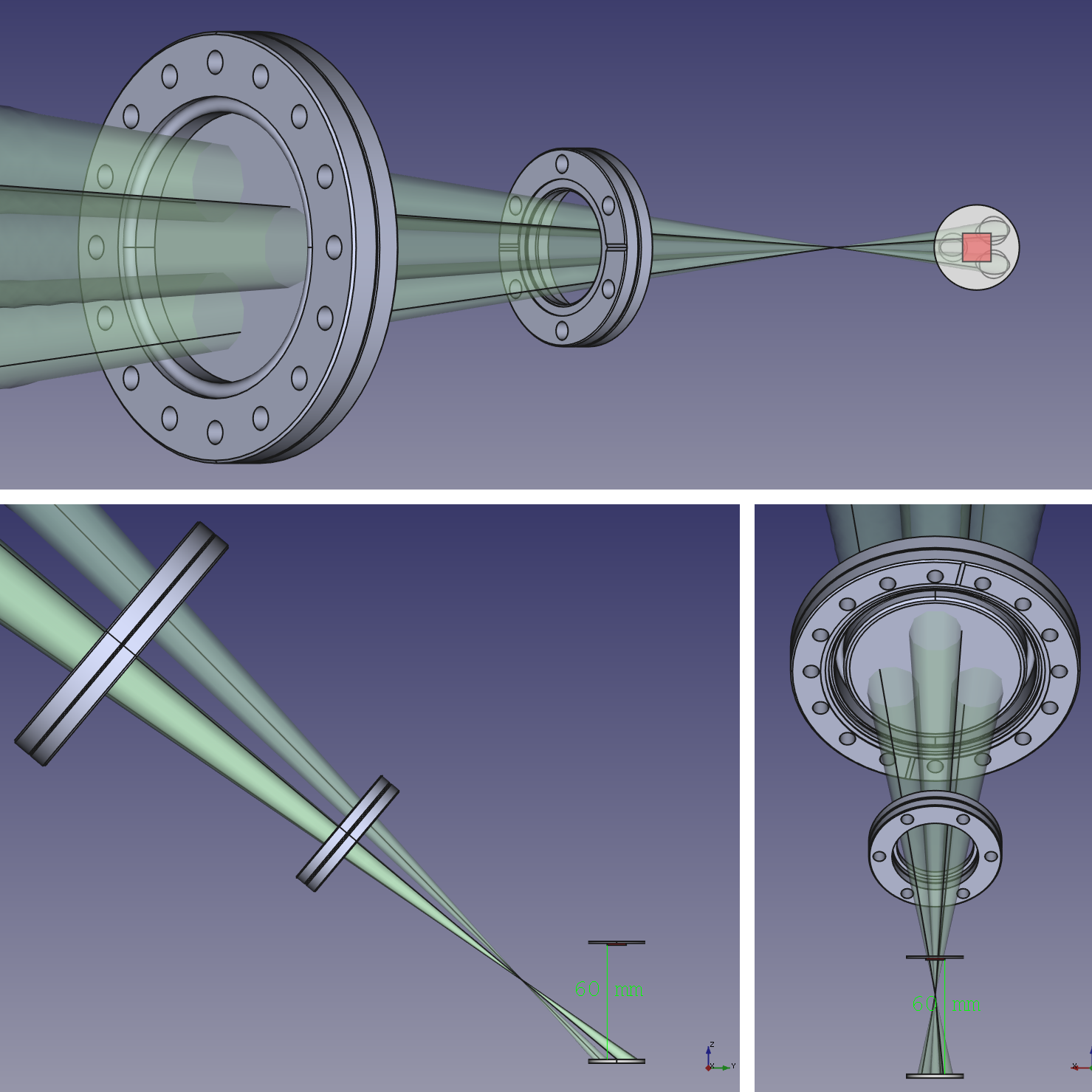}
\end{center}
\caption{Isometric views of the beam geometry.\label{labelfig_beamtargetsample234}}
\end{figure}

The sources are heated by a confocal arrangement of three fiber lasers operating at a wavelength around 1\,µm.
The hit the source material surfaces in the crucibles at an angle of 50° away from their normal, not affecting substrate handling and heating in front of the sources (Fig.~\ref{labelfig_beamtargetsample234}).
%++++++++++++++++++++++++++++++++++++++++++++++++++++++++++
%++++++++++++++++++++++++++++++++++++++++++++++++++++++++++
The lasers enter the vacuum chamber through a large quartz glass window, where their beam diameters are large enough not to damage the glass.
All three go through the same focal point and then spread again to the three source material locations.

\begin{figure}[ht]
\begin{center}
\includegraphics[width=\columnwidth]{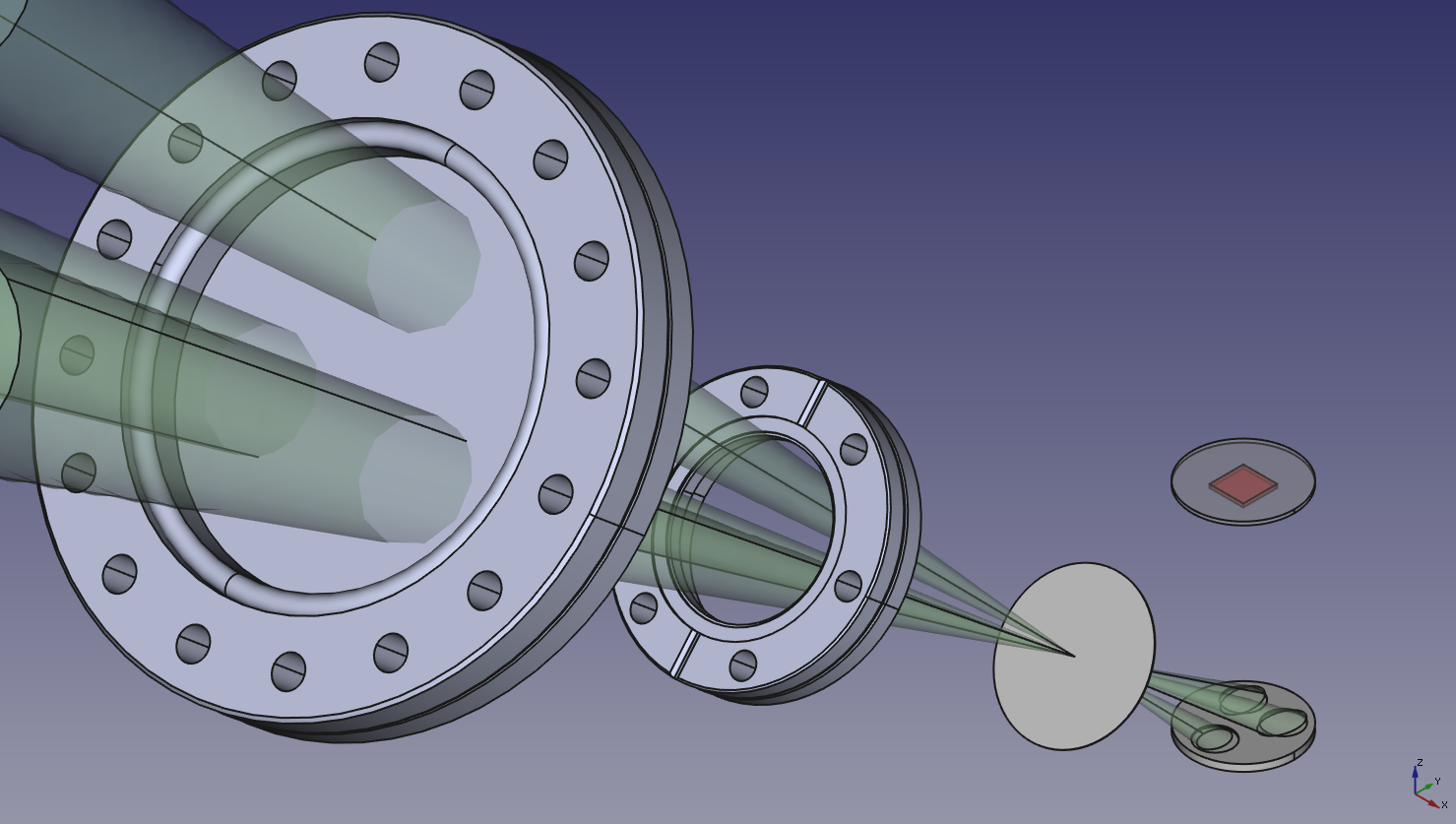}
\end{center}
\caption{Beam geometry with common aperture at focal point.\label{labelfig_beamtargetsample06}}
\end{figure}

At the common focal point, an aperture is placed to minimize coating of the laser entrance window by the material emitted from the sources (Fig.~\ref{labelfig_beamtargetsample06}).
%++++++++++++++++++++++++++++++++++++++++++++++++++++++++++
%++++++++++++++++++++++++++++++++++++++++++++++++++++++++++
Whereas the laser light is focused and can be transmitted through the aperture without loss, the material emission from the elemental sources varies as the cosine to the surface normal, and is directed into the entire half sphere in front of the sample.
The aperture thereby efficiently suppresses coating of the laser entrance windows by source material.

Shutters may be placed between the sources and the substrate, blocking the molecular beams from reaching the substrate, but allowing the laser beams to keep the sources at constant temperature.
This allows the synthesis of precisely defined layer thicknesses with steady growth conditions.

The substrate is heated by another laser heater from the back side, ideally with on-axis pyrometer temperature control.
For oxides, a long-wavelength (usually CO$_2$ at ~10\,µm) laser with 1 to 2\,kW of power is ideal.

In contrast to PLD, and as in MBE, the proposed design still operates in a pressure regime where the mean free paths are larger or just comparable to the working distances between sources and substrate.
A change in these distances by a factor of 2 or 5 does not significantly affect the deposition conditions.
The technology therefore does not have a built-in length scale such as in PLD, where the close interplay between ablation spot size, particle energy and background gas pressure lead to working distances below 100\,mm that limit the extension of the method to economic deposition on large substrates.

Being a blend of PLD and MBE approaches, the proposed deposition method is \lq amphibic\rq\ in that it can be combined with elements from both the PLD and the MBE world.
Since the deposition is thermal, and mean free paths are large enough for line-of-sight deposition, the laser-heated sources may be combined with additional, traditional MBE sources, plasma or cracked gas beam sources and the like, as long as these can be adapted to the extended pressure regime.
On the other hand, replacing one of the source heater thermal lasers by an excimer laser, one can use ablation for one or any number of the sources, while the others are operated in the thermal heating regime.
A cyclic variation of background pressures even allows a temporary transition into the native PLD mode of operation.
As the pressure limitation for the thermal sources applies only to the transport of the source material to the substrate, these may stay at operating temperature during such high pressure intervals, with their beams losing directionality and only contributing to the background gas.
They may even be used as local thermalized background gas generators in the volume between the sources and the substrate, providing high-pressure environments to decelarete particles in the PLD plume to thermal velocities for codeposition with the thermal species on the substrate.

\subsection{Discussion / Risk Assessment}
What are the problems that one may face in establishing this new technique, what challenges may have to be overcome in the actual implementation?

\subsubsection{Generality of the Approach}
As discussed in reference \onlinecite{EngelHerbert2013417}, SrTiO$_3$ is one rather extreme case or perovskite oxides in terms of high substrate temperatures and low desorption rates of volatile species.
This means that if the adsorption-controlled deposition of SrTiO$_3$ can be demonstrated, it will also be possible for many other perovskites.

\subsubsection{Validity of the Thermodynamic Extrapolation}
The validity of the thermodynamic diagrams (Figs.~\ref{labelfig_stolimited} and \ref{labelfig_oxidation}) at high temperatures can be verified by our own substrate thermination experiments.
We find that to prepare our oxide surfaces, which is basically to remove on the order of one molecular layer of SrO from the surface, we require a 200\,s anneal at 1300\,°C.
The rate of incidence (and desorption in equilibrium) of ideal gas particles on a surface is given by \cite{Suurmeijer2016}:
\begin{equation}
\frac{dN_i}{dt}=2.63\times 10^{24}\frac{p}{\sqrt{MT}} \frac{1}{m^2s}\label{labeleqn_impingementrate}
\end{equation}
with $p$ in Pa and $T$ in K, M\,=\,mass number.
From Fig.~\ref{labelfig_stolimited}, at 1300\,°C (1575\,K), the SrO equilibrium pressure is approximately 10$^{-7}$\,mbar, or 10$^{-5}$\,Pa.
With $M$\,=\,104 for SrO, the above formula yields
\begin{equation}
dN_i/dt=6.5\times10^{16}\,\mathrm{m}^{-2}\mathrm{s}^{-1}.\label{labeleqn_impingementrateresult}
\end{equation}
One surface unit cell measures approximately $a$\,=\,4\,Å$\times$4\,Å\,=\,1.6$\times$10$^{-19}$\,m$^2$.
If we need to desorb one SrO per unit cell, that makes 1/$a$\,=\,6.25$\times$10$^{18}$\,m$^{-2}$.
This value divided by the adsorption/desorption rate (\ref{labeleqn_impingementrateresult}) predicts a desorption time t\,=\,96\,s, in agreement with the experimental observations.
It thefore seems safe to trust the predictions for the location and size of the adsorption-controlled growth window for SrTiO$_3$.

\subsubsection{Laser Window Coating}
A technological problem may be the coating of the laser entrance windows that are in direct line of sight of the sources.
While the substrate heater is shielded from the sources by the sample holder, this cannot be avoided for the source heater lasers, as these need direct line of sight access to the sources.
The problem can be reduced, however, since the laser beam focus can be placed between the entrance window and the source.
The design provides an aperture at the common focal point of the three source lasers near (65\,mm before the center of) the source plate.
Due to the cosine law emission of a clean and even surface, the flux density originating from a small source surface element reaching this aperture is already cos(50°)\,=\,0.64 of the one reaching the substrate surface.
Projected via 1/r$^2$ to the laser entrance window at 332\,mm from the sources, the flux density reduces by another factor of 65$^2$/332$^2$\,=\,0.038 to a total of 0.025.
For an aperture size small compared to the source size and close to the source, this value is further reduced as an area element on the entrance window does not see the entire source surface through the aperture.
A 1\,mm$^2$ aperture, projected from a point on the entrance window to the source with a factor
\begin{equation}
\left(\frac{332+65}{332}\right)=1.43\,\mathrm{mm}^2
\end{equation}
may still be only 1/5 the size of the entire emitting surface.
The total reduction factor for the flux density reaching the laser entrance window vs.\ the substrate surface then amounts to around 0.5\,\%.
This value critically depends on the aperture size.
The lasers used for source heating, and in particular the focusing optics used, are designed for cutting sheet metal with small focal point diameters, typically well below 0.1\,mm.
We can therefore expect similar values for our setup.
We aim to fabricate the aperture in situ, e.g.\ by visually aligning the lasers in situ on the aperture plane and then letting them burn a hole through a thin Ta foil, hopefully reaching aperture areas significantly smaller than 1\,mm$^2$.
Generally, as SrTiO$_3$ is transparent at the ~1\,µm wavelength of the source heater lasers, and reflectivities of a single Fabry-Perot layer are small, we expect little influence of window coating on the performance of the source heating subsystem.

The design with a focal point between the entrance window and the source offers the additional advantage that reflected radiation is diverging (unless the source surface is concave with a shorter focal length than the beam optics), and therefore beam damage to other parts in the growth chamber as well as back-reflection into the laser optics is minimized.

\subsubsection{Gas Phase Reactions and Excitations}
The evaporated atoms of the sources pass through the laser heater beam as they travel from the source surface to the substrate.
The overlap distance is a few mm, and adsorption lines of heavier metals such as Fe, Cr and Ni all begin above 2\,µm wavelength, indicating that at the laser wavelength of ~1\,µm, the photoionization threshold is not reached, and other excitations are also absent.

At the envisioned high operating pressures, the mean free path approaches the distance between source and sample.
In an oxygen environment, oxidation of metal atoms on their way to the substrate is feasible.
This should be no problem, as long as no big oxide clusters form in the gas phase.
Experience from PLD indicates that pressures above 0.1\,mbar are necessary to form clusters, and good epitaxial growth is still possible with the deposition of such clusters on the surface.
We therefore do not expect a negative influence of gas phase reactions at the envisaged 10$^{-3}$\,mbar oxygen background pressures.
\vspace{-2mm}

\subsubsection{Source Temperatures}
\begin{figure}[ht]
\begin{center}
\includegraphics[width=\columnwidth]{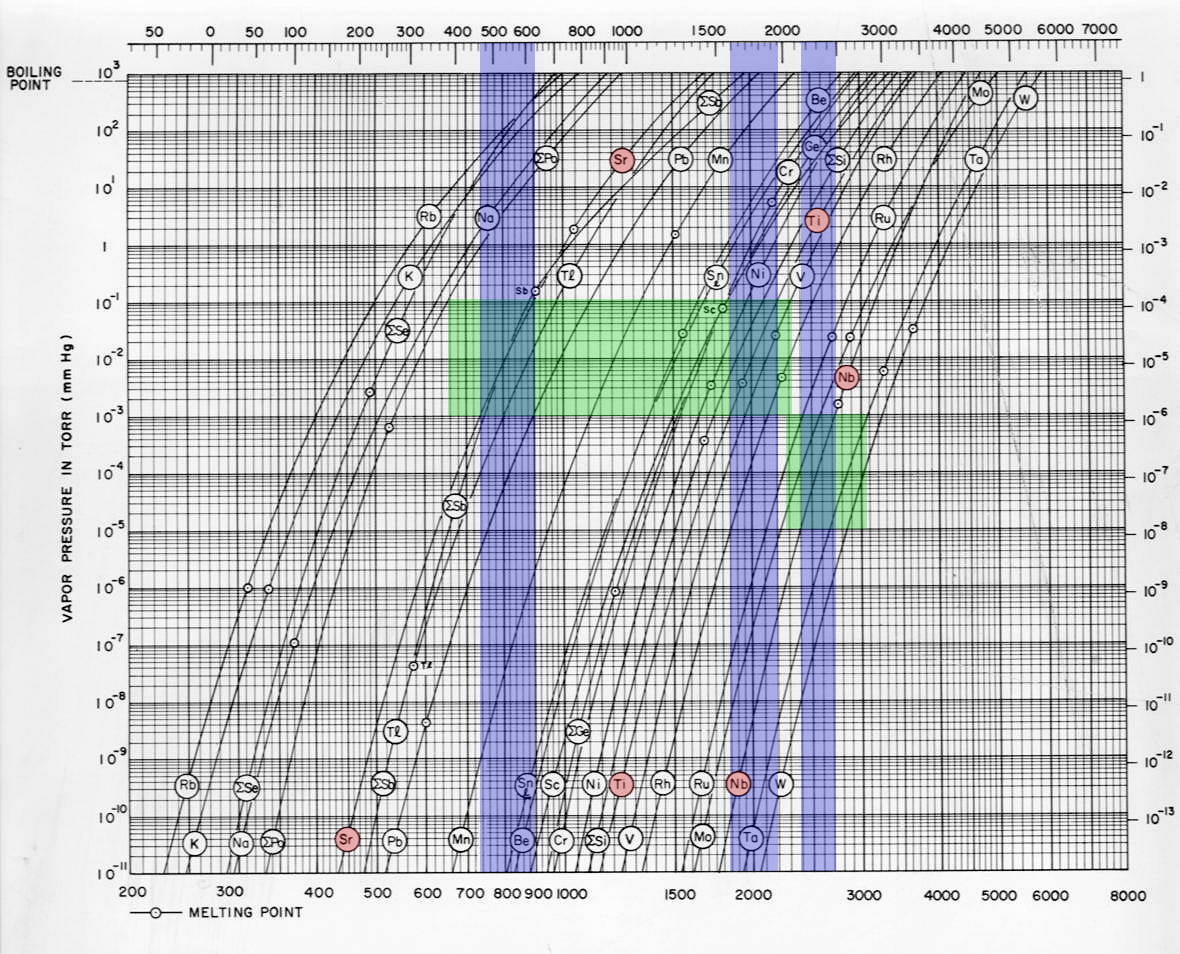}
\end{center}
\caption{Temperatures required to reach sufficient deposition fluxes for Sr, Ti and Nb.\label{labelfig_elementvaporpressures}}
\end{figure}

Figure~\ref{labelfig_elementvaporpressures} shows the vapor pressures of a number of elements, with Sr, Ti and Nb highlighted by red dots.
%++++++++++++++++++++++++++++++++++++++++++++++++++++++++++
%++++++++++++++++++++++++++++++++++++++++++++++++++++++++++
The horizontal green bars indicate typical pressures required for MBE deposition.
Given the much shorter working distance in the proposed design as compared to a standard MBE system, these are conservative estimates.
The upper range applies to deposition, pressures required for dopant concentrations are lower by two orders of magnitude.
Intersection of these ranges with the vapor pressure curves of the three elements we intend to use results in the ranges indicated by the vertical blue bars.
We arrive at 460--630\,°C for Sr, 1600--1950\,°C for Ti, and 2150--2500\,°C for Nb.
While the temperatures required for Sr deposition should be no problem, the higher temperatures for Ti, and in particular Nb, may pose a challenge.

Experience exists with the existing CO$_2$ laser heater for substrates, which is capable of heating substrates 5$\times$5$\times$0.5\,mm$^3$ to temperatures in excess of 2000\,°C with a power below 1\,kW.

E-beam evaporation of Nb is reported in the literature, I evaporated W a few years ago with a compact e-beam evaporator with a similar working distance as the one proposed here.
In our standard e-beam evaporation chamber in lab 4C11, we evaporate Ti by e-beam evaporator using the following parameters: 8\,kV, 130\,mA (1040\,W), working distance 350\,mm: growth rate 1.5\,\,Å/s.
Scaling this with the $1/r^2$ratio $350^2/60^2=34$, we can expect  one and a half orders of magnitude of headroom for the 1\,kW maximum power fiber laser proposed here.

\begin{figure}[ht]
\begin{center}
\includegraphics[width=.9\columnwidth]{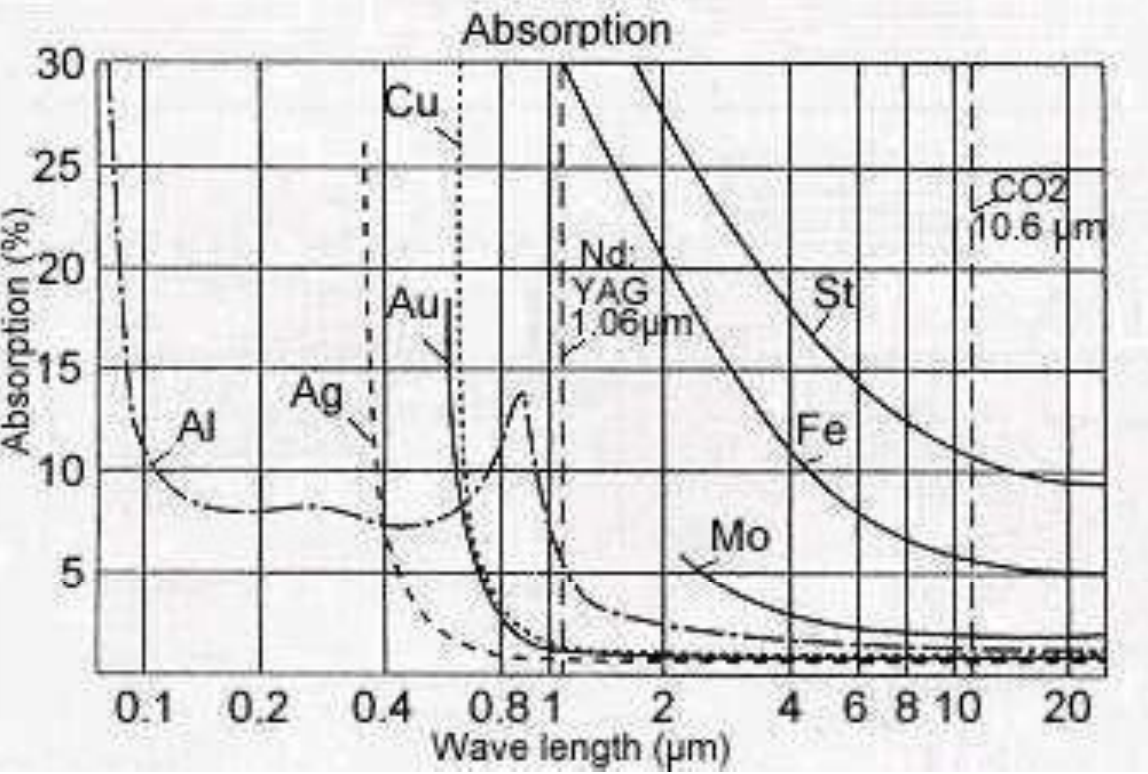}
\end{center}
\caption{Absorption of laser light (at normal incidence?) for various metals, given without reference in Ref.~\onlinecite{bergstrom2008}. \lq St\rq\ denotes steel.\label{labelfig_laserabsorptionofmetals}}
\end{figure}

All this assumes a comparable absorptance of the source material between e-beam evaporation and laser heating.
What matters for laser heating is the absorptance of the laser light at 1070\,nm wavelength, since economically feasible high-power tooling lasers all operate at this wavelength.
Data in the literature scatter a lot, as they are usually acquired from relatively rough, contaminated and oxidized surfaces around room temperature.
What we require is the absorptance at the above temperatures, usually on a liquid, very clean surface, and at an incidence angle of 50° to the surface normal.
Data exists mainly in the context of laser cutting and welding, where the surfaces are neither smooth nor clean (in particular not oxidized).
An example is shown in Fig.~\ref{labelfig_laserabsorptionofmetals}
\vspace{-2mm}
%++++++++++++++++++++++++++++++++++++++++++++++++++++++++++
%++++++++++++++++++++++++++++++++++++++++++++++++++++++++++
The general trend for metal surfaces is to absorb less and less energy with increasing wavelength.
There are strong differences between different metals, with Al, Ag, Au and Cu having low absorptance, whereas other metals such as Fe can reach values of 30 to 50\,\% at wavelengths just above 1\,µm, our targeted working point.

\begin{figure}[ht]
\begin{center}
\includegraphics[width=.87\columnwidth]{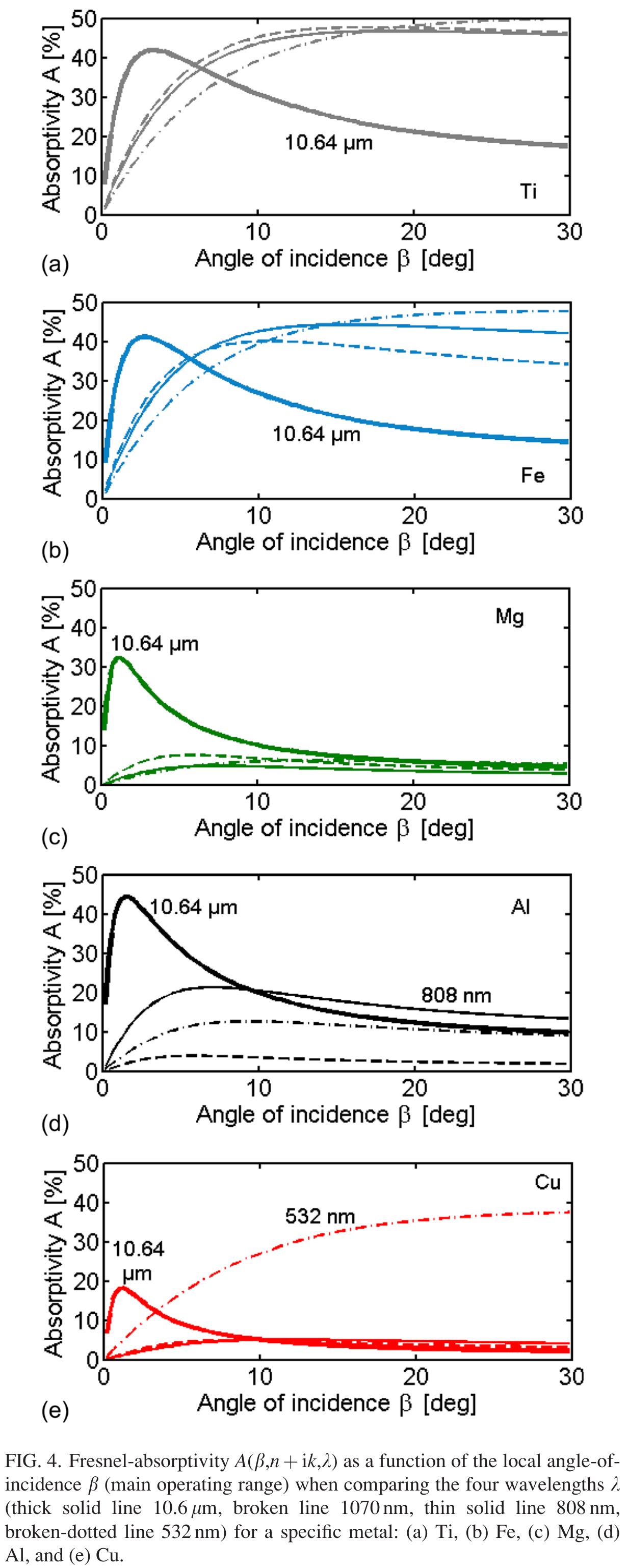}
\end{center}
\caption{Absorptance of a molten surface as a function of laser wavelength and incidence angle for various metals\cite{kaplan2014}.\label{labelfig_absorptivitymoltensurface}}
\end{figure}

I found one study looking at the absorptance of laser light by molten metal surfaces at low incidence angles ranging from 0° (parallel to the surface) to about 30°.
The important figure in our context is reproduced in Fig.~\ref{labelfig_absorptivitymoltensurface}.
%++++++++++++++++++++++++++++++++++++++++++++++++++++++++++
%++++++++++++++++++++++++++++++++++++++++++++++++++++++++++
Our angle of incidence is 40° on this scale, outside the depicted range.
Absorptance at 1070\,nm is represented by the thin solid line in each panel.
While the highly conducting metals Cu, Al and Mg have very poor absorptance, the situation is good for Fe and in particular for Ti, with absorptance values above 40\,\% at high angles.
It therefore seems feasible to achieve the required temperatures above 1600\,°C with 1\,kW of laser power.
The additional headroom of a 2\,kW laser, however, would be desirable, see Section~\ref{sect_sourceoxidation}.

I was unable to find literature values for Strontium absorptance in the relevant parameter range.
Sr seems unproblematic, however, even if we assume a low absorptance in the range of 10\,\%, as the required temperatures are low.
The main loss channel, blackbody radiation increasing as $p\propto T^4$, does not yet play a significant role at 600\,°C.
In our existing system, heating oxides with a CO$_2$ laser, we reach these temperatures with around 10\,W of power, so even assuming an absorptance of 10\,\%, we would still have one order of magnitude headroom for this source with a 1\,kW heating laser.

\begin{figure}[ht]
\begin{center}
\includegraphics[width=\columnwidth]{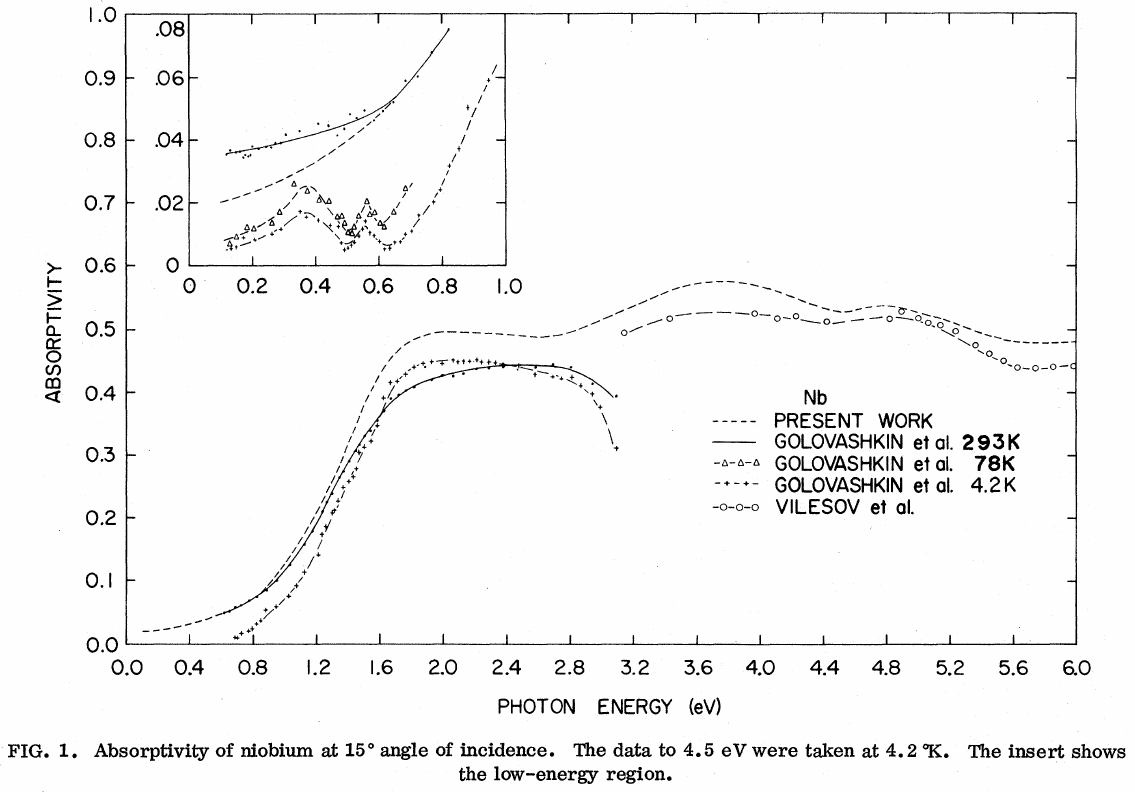}
\end{center}
\caption{Absorptivity of Niobium\cite{weaver1973}. 1070\,nm wavelength corresponds to 1.16\,eV.\label{labelfig_niobiumabsorptivity}}
\end{figure}

The situation is different for Nb, though.
Literature data indicate a low absorptance of 20\,\% at 1070\,nm Fig.~\ref{labelfig_niobiumabsorptivity}.
%++++++++++++++++++++++++++++++++++++++++++++++++++++++++++
%++++++++++++++++++++++++++++++++++++++++++++++++++++++++++
Compared to Ti, less than half the power of the laser may be coupled into the source material.
At the same time, significantly higher temperatures of 2200\,°C need to be reached.
One therefore would need at least a 2\,kW, if not a 5\,kW laser for this source, assuming a perfectly clean metallic surface (see Section~\ref{sect_sourceoxidation}.

If Nb cannot be heated to temperatures required for sufficient doping levels, other dopants may be used.
La, e.g.,\  may be used instead, which requires much lower temperatures of 1300--1600\,°C for similar vapor pressures.

In this context, it is interesting to compare the evaporation of Au with Ti in our e-beam evaporator.
For identical growth rates, where Ti requires 960\,W, evaporation of Au in the same e-beam evaporator with similar growth rate requires 1970\,W of power.
Au requires less than 1400\,°C to reach a vapor pressure of 10$^{-5}$ bar, whereas Ti reaches the same vapor pressure only at more than 1700\,°C.
At around 1500\,°C, both differ by a factor of 350 in vapor pressure, despite the doubled input power for Au.
Obviously, a similar trend of power absorption with nobility or conductivity of the metal seems to be present in e-beam evaporation.
We can conclude that for e-beam heating, the absorptivity of the electron beam is also less than one, which reduces the estimated power requirements of the laser heater.

\subsubsection{Source Oxidation\label{sect_sourceoxidation}}
In the proposed setup, we raise the oxygen partial pressures to ensure full oxidation of the growing layer and to prevent oxygen loss from the substrate crystal.
We can, therefore, expect the surfaces of the metallic sources in turn to oxidize heavily, as we deliberately work at oxygen pressures and substrate temperatures where the oxidation of the evaporated elements is possible even with molecular oxygen.
In general, this should not be a problem, as epitaxy should work both with metal atoms arriving at the surface and oxidizing there, or metal-oxygen molecules that have formed prior to arrival.
The vapor pressure of metal oxides, however, is usually distinctly higher than the one of the corresponding metal, requiring higher source temperatures to obtain the same metal flux on the substrate.

The nobler the metal, the less stable its oxide.
Reconsidering Fig.~\ref{labelfig_oxidation}, e.g., we see that Sr no longer reacts with 10$^{-3}$\,mbar molecular oxygen at temperatures above 1500\,°C.
For the case of Sr, we can therefore expect to decompose an oxide layer on the source surface, or sublimate/evaporate it with the 1\,kW of power planned for this source.

\begin{figure}[ht]
\begin{center}
\includegraphics[width=\columnwidth]{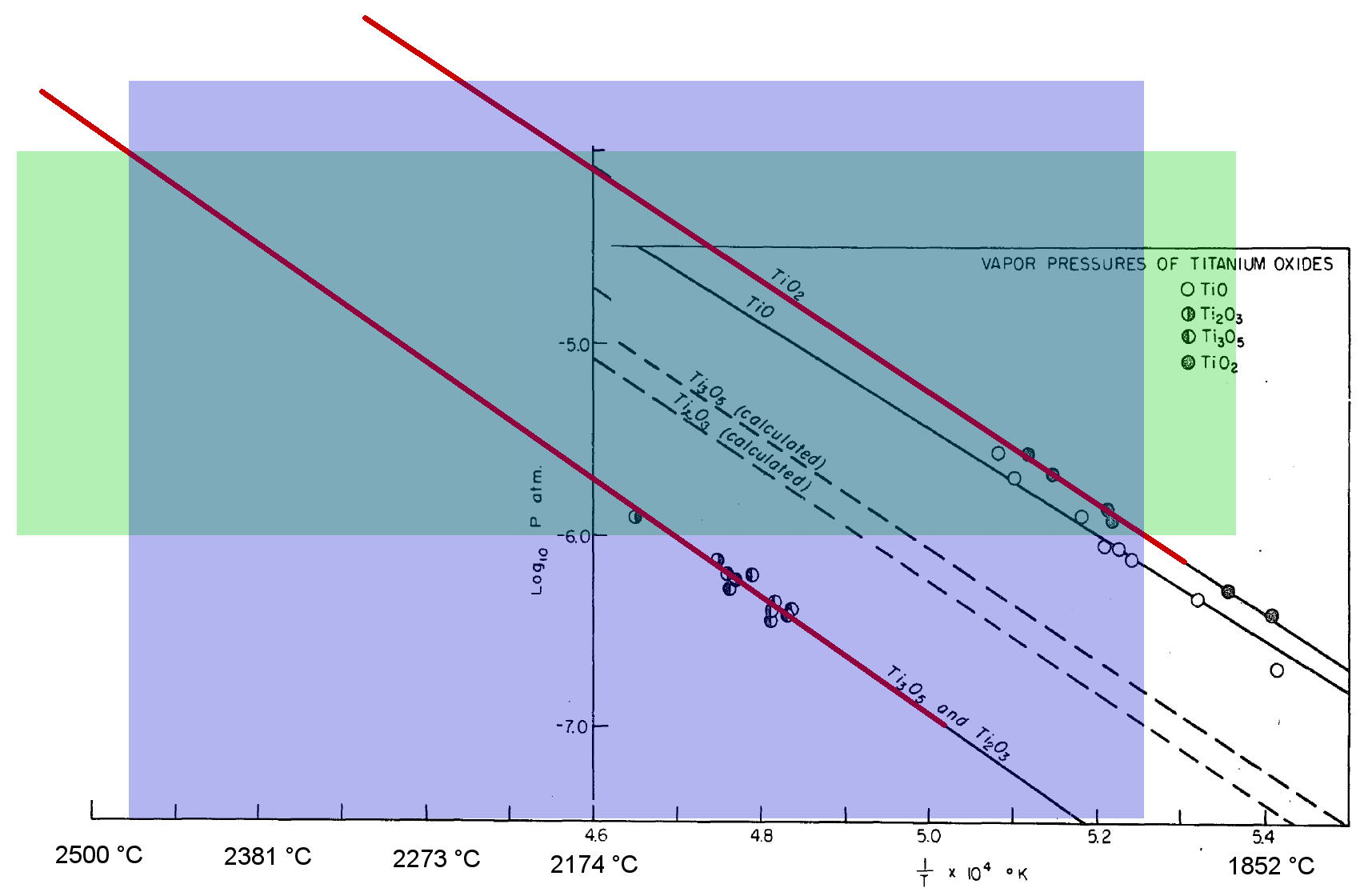}
\end{center}
\caption{Vapor pressures of various Ti oxides \cite{groves1954}.\label{labelfig_titaniumoxidesvaporpressures}}
\end{figure}

Ti, however, forms a much more stable oxide.
The vapor pressure of Ti oxides spans a large range (Fig.~\ref{labelfig_titaniumoxidesvaporpressures}).
\vspace{-2mm}
%++++++++++++++++++++++++++++++++++++++++++++++++++++++++++
%++++++++++++++++++++++++++++++++++++++++++++++++++++++++++
In this figure, the green bar shows the required vapor pressure for deposition as in Fig.~\ref{labelfig_elementvaporpressures}.
The required temperature range, again indicated in blue, spans temperatures from 1900\,°C to almost 2500\,°C.
If all surface Ti would form oxides, in particular with lower oxidation states, this would present a substantial challenge.

On the other hand, Ti effusion cells are available with operating temperatures up to 2000\,°C, and the the thermal evaporation of Ti in oxide MBE has been demonstrated using a resistively heated, hollow Ti ball\cite{schlom1996}.
This source was able to operate in background pressures of ozone up to 5$\times$10$^{-5}$\,Torr, just one and a half orders of magnitude below the 10$^{-3}$ mbar of molecular oxygen we are targeting.
Given that ozone is usually orders of magnitude more reactive than molecular oxygen (Fig.~\ref{labelfig_oxidation}), this may allow stable operation of the Ti source with at most a thin layer of oxide on the surface.

\begin{figure}[ht]
\begin{center}
\includegraphics[width=.7\columnwidth]{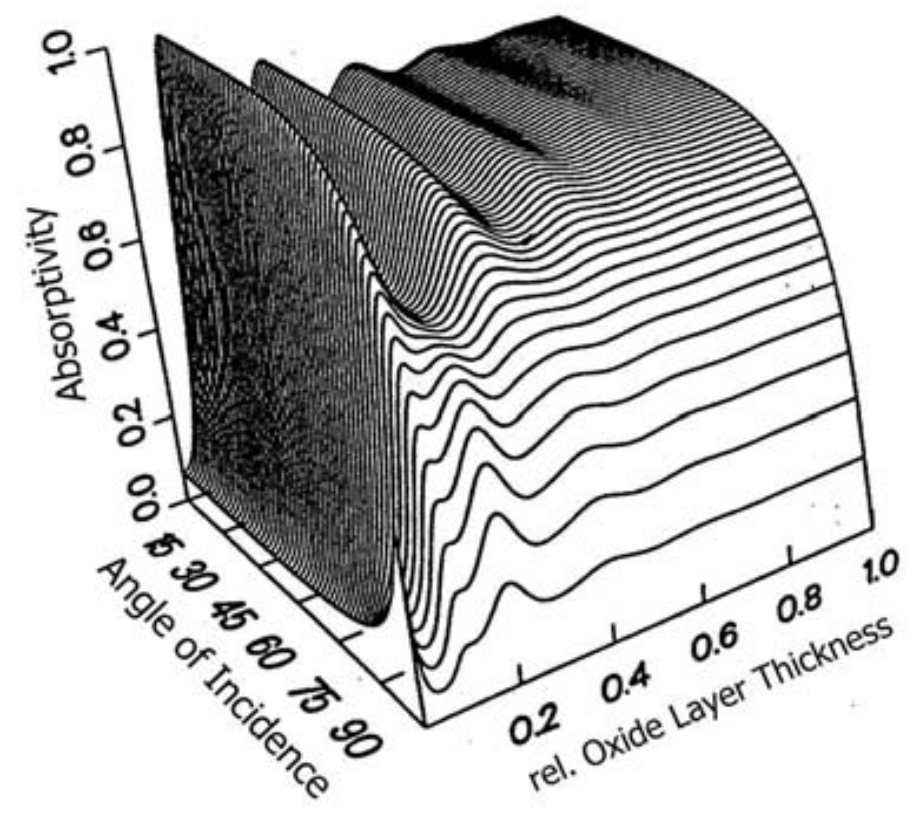}
\end{center}
\caption{Absorptance of a metal surface coated by an oxide layer, as a function of layer thickness and incidence angle\cite{franke1994}.\label{labelfig_absorptivitywithoxidelayer}}
\end{figure}

Simulations of laser absorption of metal surfaces coated by an oxide layer indicate a dramatic increase of the absorptance, in particular for non-normal incidence as is the case in our setup (Fig.~\ref{labelfig_absorptivitywithoxidelayer}).
%++++++++++++++++++++++++++++++++++++++++++++++++++++++++++
%++++++++++++++++++++++++++++++++++++++++++++++++++++++++++
This may help to compensate the vapor pressure gap between metals and their oxides by increasing the power available to reach higher temperatures.

\subsubsection{Source Purity}
Source purity should not be a concern for the metal sources, as elemental metals are usually available with higher purities than the ceramic targets mostly used in PLD.
Care needs to be taken to ensure a supply of clean oxygen, since at an oxygen pressure of 10$^{-3}$\,mbar, a partial pressure of contaminants in the oxygen to match the base pressure of an MBE type chamber of 10$^{-10}$\,mbar or better, would require an oxygen purity of 10$^{-7}$.

\subsubsection{Temperature Control}
As in MBE, thermocouple temperature sensors may be placed at the back of each of the source material crucibles, allowing a direct measurement of the temperatures at this position.
Since the laser heats only the front surface of the source material, the temperature is not uniform around the crucible, and significant spatial and temporal gradients may exist.
On the other hand, the sources will be operated at high temperatures, where the dominant loss channel are radiative losses.
These can be minimized around the surfaces of the source crucibles not facing the substrate by providing multiple layers of metal foil shielding.
Strong radiative coupling and reflections will then lead to a more uniform temperature distribution which will hopefully allow us to measure rather accurate and reproducible source temperatures.
As the laser power is usually stable to within 1\,\%, even a pure power control, without temperature feedback, should allow the growth in the self-adjusting growth window, where deviations of a few percent from the ideal flux ratios should make no difference for the film stoichiometry.

\subsubsection{process characterization}
To characterize the growth, a high-pressure RHEED system allows the characterization of the surface configuration and, if present, growth oscillations.
For the analysis of desorption, a cross-beam quadrupole mass spectrometer may be attached to one of the source ports of the chamber, with a snout reaching in close to the sample, to analyze desorbing species from the surface.
Experience at the synchrotron machine has shown that this works\cite{perumal2014}, and the detection of desorbing SrO would be direct proof of the desorption-controlled growth mode.
The essential parts for this quadrupole measurement addon (mass spectrometer, turbo pump, roughing pump, flanges) have already been purchased and would be available for the project.

\section{Demonstrator}
I propose to build a prototype laser epitaxy system capable of growing doped SrTiO$_3$ by implementing three independent laser-heated elemental sources.
This system will be used to test the above hypothesis that adsorption-controlled growth of SrTiO$_3$ is possible in such an environment.

By placing the growth chamber next to the existing beamline of one of the CO$_2$ heating lasers, the substrate may be heated by adding another branch to one of the beamlines and using the existing laser.
This beamline branch would also be used to test a new, simpler beamline configuration, with a conical mirror instead of the pi shaper used in the existing branches.
The Bragg mirror and control pyrometer, however, need to be duplicated as they are located at the end of the branch, directly in front of the entrance window into the chamber.
The new branch will be connected via a pneumatic mirror that may be switched to direct the beam into either one of the connected beamlines.

\begin{figure}[ht]
\begin{center}
\includegraphics[width=\columnwidth]{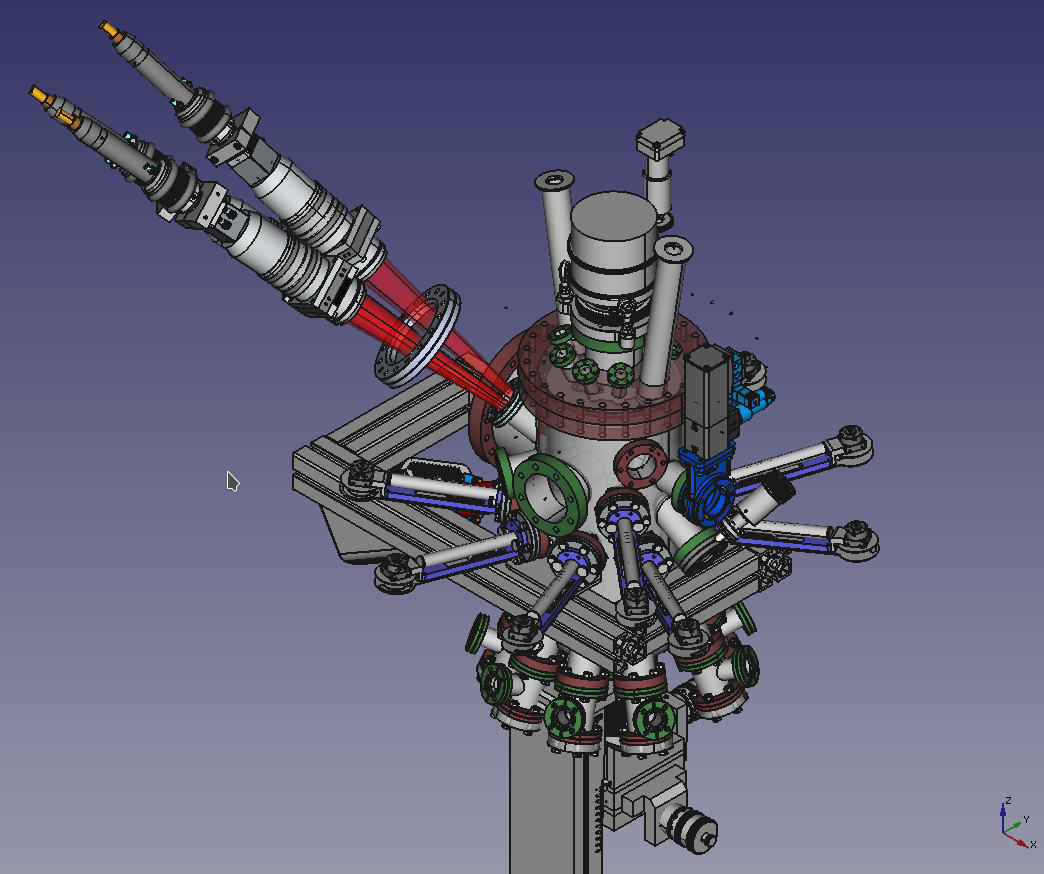}
\end{center}
\caption{Beam geometry of triple laser source heater on mini\-MBE system.\label{labelfig_conesminimbe}}
\end{figure}

\begin{figure}[ht]
\begin{center}
\includegraphics[width=\columnwidth]{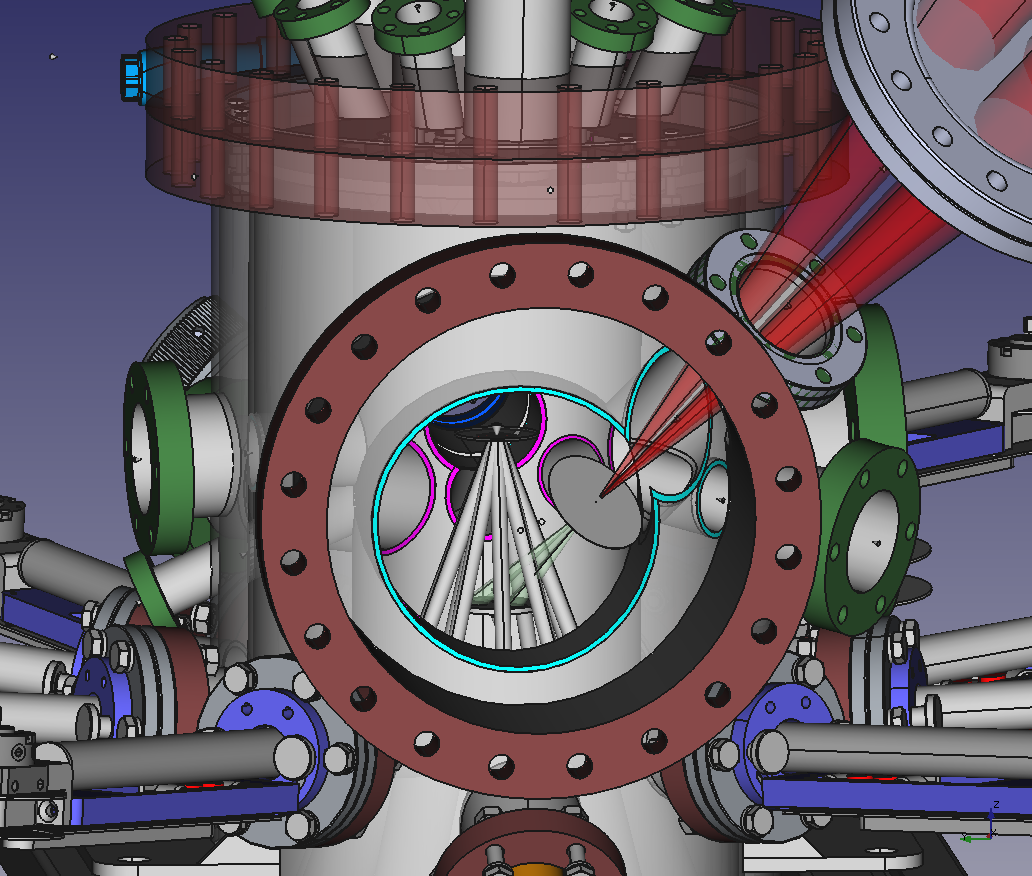}
\end{center}
\caption{Beam geometry of triple laser incidence on target disc, together with conventional MBE sources.\label{labelfig_conesontarget}}
\end{figure}

\begin{figure}[ht]
\begin{center}
\includegraphics[width=\columnwidth]{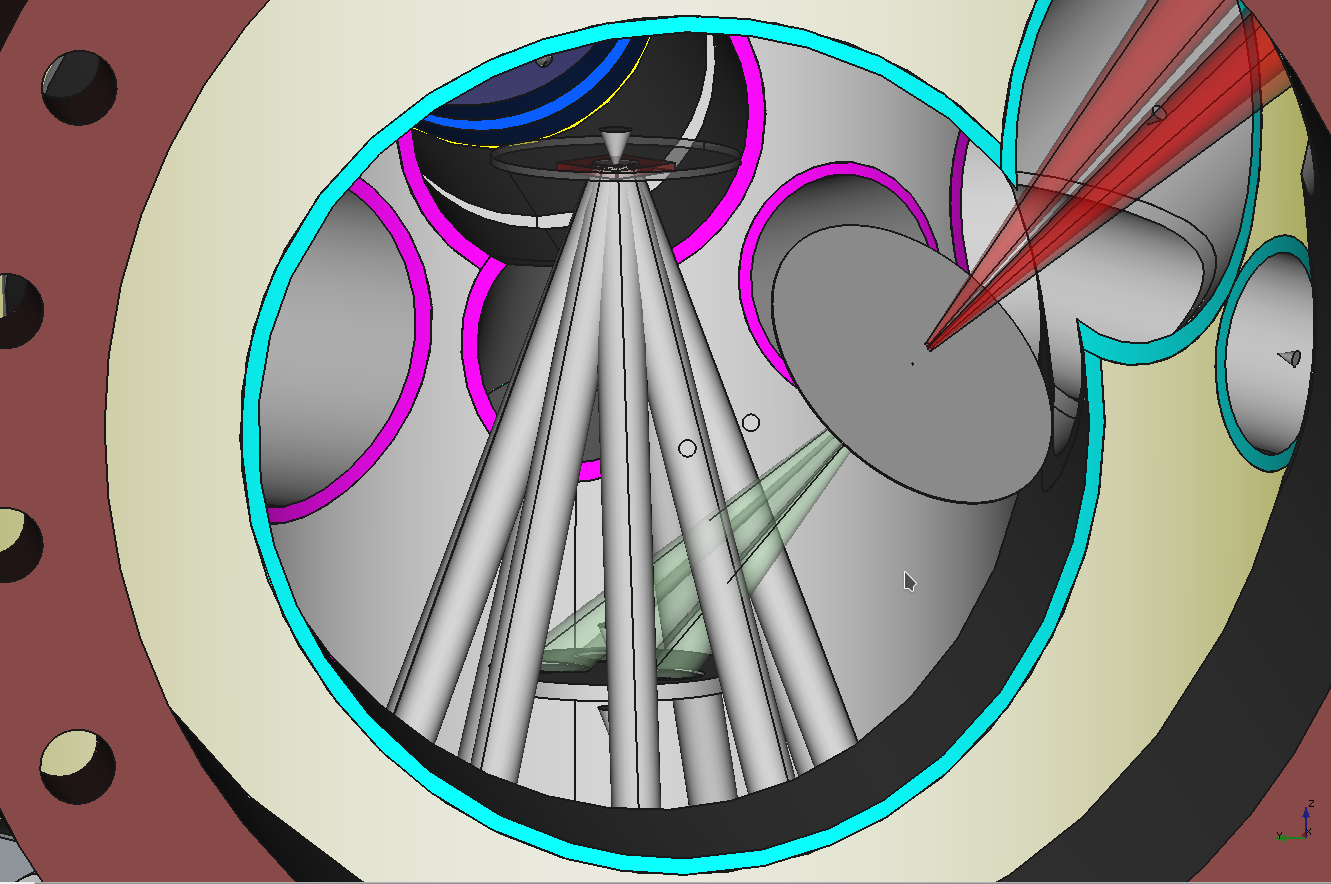}
\end{center}
\caption{Geometry of target, sample, sources and laser beams.\label{labelfig_laserheads3}}
\end{figure}

For the growth chamber, the existing miniMBE chamber may be used, as it already provides a port for optical access to the required location (Fig.~\ref{labelfig_conesminimbe}).
%++++++++++++++++++++++++++++++++++++++++++++++++++++++++++
%++++++++++++++++++++++++++++++++++++++++++++++++++++++++++
As shown in Fig.~\ref{labelfig_beamtargetsample06}, a conical extender will have to be mounted to provide a DN\,100\,CF entrance window, large enough and close enough to the laser focusing optics so that no beam damage occurs at the window.
When going beyond 1\,kW per laser, we even may have to go to a DN\,160\,CF size window.
This unit may be designed to also include a holder for the focal plane aperture (Figs.~\ref{labelfig_conesontarget}, \ref{labelfig_laserheads3}) so that preliminary alignment of the beams is possible ex situ, before mounting the unit on the chamber.

A closeup view of the sample--source geometry is shown in Fig.~\ref{labelfig_conesontarget}, demonstrating that the setup may even be combined with fluxes from the standard MBE sources provided by the chamber geometry.
%++++++++++++++++++++++++++++++++++++++++++++++++++++++++++
%++++++++++++++++++++++++++++++++++++++++++++++++++++++++++
%++++++++++++++++++++++++++++++++++++++++++++++++++++++++++
%++++++++++++++++++++++++++++++++++++++++++++++++++++++++++

A new, differentially pumped high-pressure RHEED gun is becoming available that is very compact in size, fitting the size of the test chamber.
We envisage using and testing this design to characterize the growth.

A dedicated chamber, without additional effusion sources, could be significantly smaller, with ports for
\begin{itemize}
\item substrate holder
\item target holder
\item sample transfer
\item viewport
\item thermal laser port ($\times$2 at 180° around the vertical for 6 sources)
\item RHEED gun
\item RHEED screen
\end{itemize}
and a shroud for LN$_2$ or water cooling.
A chamber diameter with standard size  DN\,160\,CF seems feasible.
An additional ablation laser port at 90° to the thermal ports would allow use as a PLD system.

If possible, a differentially pumped quadrupole mass spectrometer should be envisaged to study the species desorbing from the substrate.
This would allow for a detailed test of the thermodynamic estimates that underly the project and could significantly enhance the scientific output and validity of the study.
The main parts for such an analysis device (cross-beam QMS and turbopump for differential pumping) have already been ordered and would be available for this purpose.

\noindent To realize the project, we need the following components:

\medskip

\noindent\begin{tabular}{l p{6.8cm} r}
1 & laser unit 1\,kW TruDisk 1000 &  \censor{86}\,k€ \\
1 & laser unit 2\,kW TruDisk 2000 &  \censor{117}\,k€ \\
1 & laser unit 3\,kW TruDisk 3000 &  \censor{126}\,k€ \\
3 & fiber end optics $\approx$400\,mm focal length & 3$\times$\censor{3.4}\,k€ \\
1 & beam switch with optical bench for beam shaping and Bragg mirror for existing CO$_2$ laser & \censor{100}\,k€\\
1 & pyrometer for temperature control of CO$_\mathrm{2}$ laser & \censor{14}\,k€ \\
1 & ozone source with flow control, lines and pumping & \censor{55}\,k€ \\
1 & high pressure RHEED system & \censor{32.5}\,k€ \\
\hline \\
 & \textbf{Total} & \censor{540.7}\,k€ \\
\end{tabular}

\medskip

\bibliography{cld.bib}

\begin{thebibliography}{10}%
\makeatletter
\providecommand \@ifxundefined [1]{%
 \@ifx{#1\undefined}
}%
\providecommand \@ifnum [1]{%
 \ifnum #1\expandafter \@firstoftwo
 \else \expandafter \@secondoftwo
 \fi
}%
\providecommand \@ifx [1]{%
 \ifx #1\expandafter \@firstoftwo
 \else \expandafter \@secondoftwo
 \fi
}%
\providecommand \natexlab [1]{#1}%
\providecommand \enquote  [1]{``#1''}%
\providecommand \bibnamefont  [1]{#1}%
\providecommand \bibfnamefont [1]{#1}%
\providecommand \citenamefont [1]{#1}%
\providecommand \href@noop [0]{\@secondoftwo}%
\providecommand \href [0]{\begingroup \@sanitize@url \@href}%
\providecommand \@href[1]{\@@startlink{#1}\@@href}%
\providecommand \@@href[1]{\endgroup#1\@@endlink}%
\providecommand \@sanitize@url [0]{\catcode `\\12\catcode `\$12\catcode
  `\&12\catcode `\#12\catcode `\^12\catcode `\_12\catcode `\%12\relax}%
\providecommand \@@startlink[1]{}%
\providecommand \@@endlink[0]{}%
\providecommand \url  [0]{\begingroup\@sanitize@url \@url }%
\providecommand \@url [1]{\endgroup\@href {#1}{\urlprefix }}%
\providecommand \urlprefix  [0]{URL }%
\providecommand \Eprint [0]{\href }%
\providecommand \doibase [0]{https://doi.org/}%
\providecommand \selectlanguage [0]{\@gobble}%
\providecommand \bibinfo  [0]{\@secondoftwo}%
\providecommand \bibfield  [0]{\@secondoftwo}%
\providecommand \translation [1]{[#1]}%
\providecommand \BibitemOpen [0]{}%
\providecommand \bibitemStop [0]{}%
\providecommand \bibitemNoStop [0]{.\EOS\space}%
\providecommand \EOS [0]{\spacefactor3000\relax}%
\providecommand \BibitemShut  [1]{\csname bibitem#1\endcsname}%
\let\auto@bib@innerbib\@empty
%</preamble>
\bibitem [{\citenamefont {Jousten}(2008)}]{Jousten2008}%
  \BibitemOpen
  \bibinfo {editor} {\bibfnamefont {K.}~\bibnamefont {Jousten}},\ ed.,\ \href
  {http://www.amazon.com/Handbook-Vacuum-Technology-Karl-Jousten/dp/3527407235}
  {\emph {\bibinfo {title} {Handbook of Vacuum Technology}}}\ (\bibinfo
  {publisher} {Wiley-VCH},\ \bibinfo {address} {Weinheim},\ \bibinfo {year}
  {2008})\BibitemShut {NoStop}%
\bibitem [{\citenamefont {Engel-Herbert}(2013)}]{EngelHerbert2013417}%
  \BibitemOpen
  \bibfield  {author} {\bibinfo {author} {\bibfnamefont {R.}~\bibnamefont
  {Engel-Herbert}},\ }\bibfield  {title} {\bibinfo {title} {Chapter 17 -
  Molecular beam epitaxy of complex oxides},\ }in\ \href
  {https://doi.org/https://doi.org/10.1016/B978-0-12-387839-7.00017-8} {\emph
  {\bibinfo {booktitle} {Molecular Beam Epitaxy}}},\ \bibinfo {editor} {edited
  by\ \bibinfo {editor} {\bibfnamefont {M.}~\bibnamefont {Henini}}}\ (\bibinfo
  {publisher} {Elsevier},\ \bibinfo {address} {Oxford},\ \bibinfo {year}
  {2013})\ pp.\ \bibinfo {pages} {417 -- 449}\BibitemShut {NoStop}%
\bibitem [{\citenamefont {Suurmeijer}\ \emph {et~al.}(2016)\citenamefont
  {Suurmeijer}, \citenamefont {Mulder},\ and\ \citenamefont
  {Verhoeven}}]{Suurmeijer2016}%
  \BibitemOpen
  \bibfield  {author} {\bibinfo {author} {\bibfnamefont {B.}~\bibnamefont
  {Suurmeijer}}, \bibinfo {author} {\bibfnamefont {T.}~\bibnamefont {Mulder}},\
  and\ \bibinfo {author} {\bibfnamefont {J.}~\bibnamefont {Verhoeven}},\ }\href
  {http://www.book-vacuum-science-technology.com} {\emph {\bibinfo {title}
  {Vacuum Science and Technology}}}\ (\bibinfo  {publisher} {The High Tech
  Institute and Settels Savenije van Amelsvoort},\ \bibinfo {address}
  {Eindhoven},\ \bibinfo {year} {2016})\BibitemShut {NoStop}%
\bibitem [{\citenamefont {Bergstr\"om}(1994)}]{bergstrom2008}%
  \BibitemOpen
  \bibfield  {author} {\bibinfo {author} {\bibfnamefont {D.}~\bibnamefont
  {Bergstr\"om}},\ }\href@noop {} {\bibinfo {title} {The absorption of laser
  light by rough metal surfaces}} (\bibinfo {year} {1994})\BibitemShut
  {NoStop}%
\bibitem [{\citenamefont {Kaplan}(2014)}]{kaplan2014}%
  \BibitemOpen
  \bibfield  {author} {\bibinfo {author} {\bibfnamefont {A.~F.~H.}\
  \bibnamefont {Kaplan}},\ }\bibfield  {title} {\bibinfo {title} {Laser
  absorptivity on wavy molten metal surfaces: Categorization of different
  metals and wavelengths},\ }\href {https://doi.org/10.2351/1.4833936}
  {\bibfield  {journal} {\bibinfo  {journal} {Journal of Laser Applications}\
  }\textbf {\bibinfo {volume} {26}},\ \bibinfo {pages} {012007} (\bibinfo
  {year} {2014})},\ \Eprint
  {https://arxiv.org/abs/https://doi.org/10.2351/1.4833936}
  {https://doi.org/10.2351/1.4833936} \BibitemShut {NoStop}%
\bibitem [{\citenamefont {Weaver}\ \emph {et~al.}(1973)\citenamefont {Weaver},
  \citenamefont {Lynch},\ and\ \citenamefont {Olson}}]{weaver1973}%
  \BibitemOpen
  \bibfield  {author} {\bibinfo {author} {\bibfnamefont {J.~H.}\ \bibnamefont
  {Weaver}}, \bibinfo {author} {\bibfnamefont {D.~W.}\ \bibnamefont {Lynch}},\
  and\ \bibinfo {author} {\bibfnamefont {C.~G.}\ \bibnamefont {Olson}},\
  }\bibfield  {title} {\bibinfo {title} {Optical properties of niobium from 0.1
  to 36.4\,eV},\ }\href {https://doi.org/10.1103/PhysRevB.7.4311} {\bibfield
  {journal} {\bibinfo  {journal} {Phys. Rev. B}\ }\textbf {\bibinfo {volume}
  {7}},\ \bibinfo {pages} {4311} (\bibinfo {year} {1973})}\BibitemShut
  {NoStop}%
\bibitem [{\citenamefont {Groves}(1954)}]{groves1954}%
  \BibitemOpen
  \bibfield  {author} {\bibinfo {author} {\bibfnamefont {W.~O.}\ \bibnamefont
  {Groves}},\ }\href@noop {} {\bibinfo {title} {Vapor-solid equilibria in the
  titanium-oxygen system}} (\bibinfo {year} {1954})\BibitemShut {NoStop}%
\bibitem [{\citenamefont {Theis}\ and\ \citenamefont
  {Schlom}(1996)}]{schlom1996}%
  \BibitemOpen
  \bibfield  {author} {\bibinfo {author} {\bibfnamefont {C.~D.}\ \bibnamefont
  {Theis}}\ and\ \bibinfo {author} {\bibfnamefont {D.~G.}\ \bibnamefont
  {Schlom}},\ }\bibfield  {title} {\bibinfo {title} {Cheap and stable titanium
  source for use in oxide molecular beam epitaxy systems},\ }\href
  {https://doi.org/10.1116/1.580185} {\bibfield  {journal} {\bibinfo  {journal}
  {Journal of Vacuum Science \& Technology A: Vacuum, Surfaces, and Films}\
  }\textbf {\bibinfo {volume} {14}},\ \bibinfo {pages} {2677} (\bibinfo {year}
  {1996})},\ \Eprint {https://arxiv.org/abs/https://doi.org/10.1116/1.580185}
  {https://doi.org/10.1116/1.580185} \BibitemShut {NoStop}%
\bibitem [{\citenamefont {Franke}(1994)}]{franke1994}%
  \BibitemOpen
  \bibfield  {author} {\bibinfo {author} {\bibfnamefont {J.~W.}\ \bibnamefont
  {Franke}},\ }\href@noop {} {\bibinfo {title} {Modellierung und Optimierung
  des Laserstrahlschneidens niedriglegierter Stähle}} (\bibinfo {year}
  {1994})\BibitemShut {NoStop}%
\bibitem [{\citenamefont {Perumal}\ \emph {et~al.}(2014)\citenamefont
  {Perumal}, \citenamefont {Braun}, \citenamefont {Riechert},\ and\
  \citenamefont {Calarco}}]{perumal2014}%
  \BibitemOpen
  \bibfield  {author} {\bibinfo {author} {\bibfnamefont {K.}~\bibnamefont
  {Perumal}}, \bibinfo {author} {\bibfnamefont {W.}~\bibnamefont {Braun}},
  \bibinfo {author} {\bibfnamefont {H.}~\bibnamefont {Riechert}},\ and\
  \bibinfo {author} {\bibfnamefont {R.}~\bibnamefont {Calarco}},\ }\bibfield
  {title} {\bibinfo {title} {Growth control of epitaxial GeTe–Sb$_2$Te$_3$ films
  using a line-of-sight quadrupole mass spectrometer},\ }\href
  {https://doi.org/https://doi.org/10.1016/j.jcrysgro.2014.03.039} {\bibfield
  {journal} {\bibinfo  {journal} {Journal of Crystal Growth}\ }\textbf
  {\bibinfo {volume} {396}},\ \bibinfo {pages} {50 } (\bibinfo {year}
  {2014})}\BibitemShut {NoStop}%
\end{thebibliography}%

\end{document}